\documentclass[12pt,a4paper]{article}

\usepackage{amsmath}
\usepackage{amsfonts}
\usepackage{amssymb}
\usepackage{graphicx}
\usepackage{graphics}
\usepackage{tikz}
\definecolor{green}{cmyk}{0.5,0.0,0.6,0.3}
\usepackage{authblk}

\usepackage{harvard}
\usepackage{lineno}
\usepackage{setspace}

\begin{document}
\title{Persistence of complex food webs in metacommunities}

\author[1]{Gesa A. B\"ohme \thanks{gesa@pks.mpg.de}}
\affil[1]{Max-Planck Institute for the Physics of Complex Systems, Dresden, Germany}
\author[2]{Thilo Gross \thanks{thilo.gross@physics.org}}
\affil[2]{University of Bristol, Department of Engineering Mathematics, Bristol, UK}

\date{}

\maketitle{}



\begin{abstract}
Metacommunity theory is considered a promising approach for explaining species diversity and food web complexity.
Recently Pillai et al.\ proposed a simple modeling framework for the dynamics of food webs at the metacommunity level. 
Here, we employ this framework to compute general conditions for the persistence of complex food webs in metacommunities.
The persistence conditions found depend on the connectivity of the resource patches and the structure of the assembled food web,
thus linking the underlying spatial patch-network and the species interaction network.
We find that the persistence of omnivores is more likely when it is feeding on (a) prey on low trophic levels, and (b) prey on similar trophic levels.
\end{abstract}

\newpage
\section{Introduction}
A central aim of ecology is to understand the emergence and maintenance of the enormous diversity of ecological species.
Both in aquatic and terrestrial environments the number of niches created by primary abiotic factors is relatively limited.
Most ecological communities thus exhibit a self-sustaining diversity where species present in the system open up niches for others.

Perhaps the simplest example for biotically created niches is found in predator-prey interactions. 
By persisting in a given system a species can open up a niche for a predator feeding on that species.  
This mechanism can in principle enable the coexistence of a large numbers of species in complex food webs. 

\begin{figure}
\begin{center}
\includegraphics[ width=.75\textwidth]{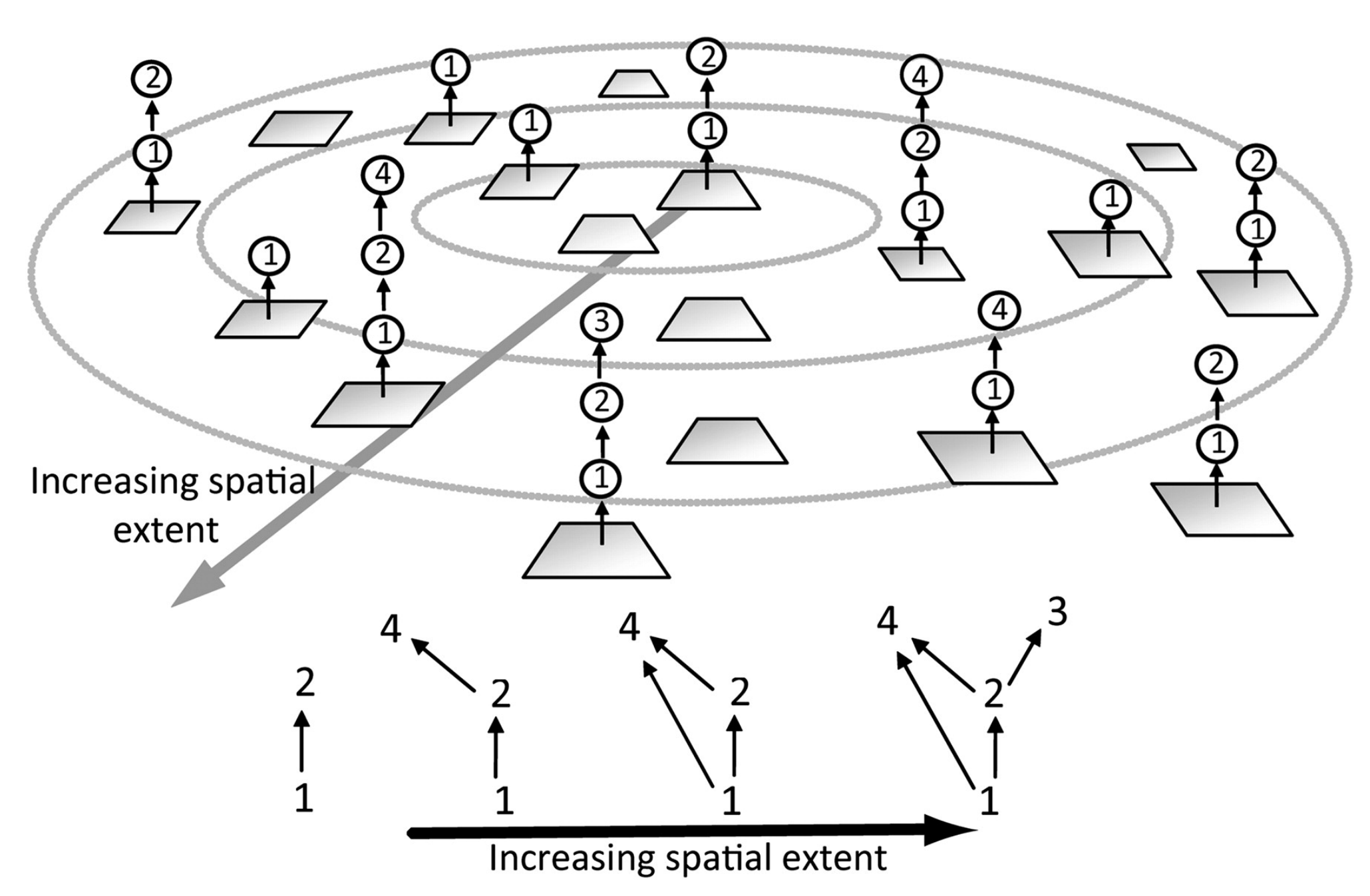}
\caption{Schematic illustration of the emergence of complex food webs. Spatially coupled local patches sustaining food chains give rise to complex food webs on the landscape scale. Figure reprinted from \protect\cite{pillai2011} .}
\label{figPillai}
\end{center}
\end{figure}

Recent investigations have identified several factors that contribute to the stability of large food webs and thus promote diversity \cite{mccann1998,brose2006,vandermeer2006,gross2009}.
However, growing evidence suggests that, at least in some systems, the food web emerges only on a regional scale, whereas simple food chains are observed if specific locations (patches) are considered in isolation (see Fig.~\ref{figPillai})\cite{pillai2011}. 
This points to a need for a meta-community perspective, in which one explicitly accounts for the dispersal of species across a network of patches. 
In particular, one can ask if and how different communities can be sustained in a network of similar patches. 

The dynamics of dispersal in a system of multiple patches was first discussed in the context of metapopulation theory, introduced by Levins \cite{levins1969,hanski1998}.
The simplest metapopulation model considers only a single species that spreads over a system of identical patches, such that each patch is either occupied by the species or vacant. 
Extending this framework to allow for multiple populations per patch leads to metacommunities.

In the present paper we focus on a class of metacommunity models that has recently been proposed by Pillai et al.\ 
\cite{pillai2010,pillai2011,pillai2012}.
These models are relatively simple and therefore conducive to a detailed analysis.  
At their center stands the assumption that within each patch competitive exclusion precludes the formation of complex food webs. 
Thus any single patch can only harbor a simple food chain. 
However, as different food chains can be realized in different patches a complex food web emerges at the landscape level (see Fig.~\ref{figPillai}). 

Pillai et al.~discuss the conditions for specific examples, when complex food webs can persist indefinitely in the metacommunity. Here we continue this work and show that the computation of persistence conditions follows a general pattern. 
Exploiting this pattern, we present a general approach allowing the direct computation of conditions for the persistence of an arbitrary given food web in the metacommunity. 

We start in Sec.~\ref{MetaCommTheory} with a detailed description of the modeling framework. 
In Sec.~\ref{ChainTop}, we compute conditions for food chain persistence, introducing a concept which is then used to compute persistence conditions for more complex food webs in the subsequent sections.
Based on Sec.~\ref{Omnivores}, where we consider a simple example involving omnivory, we formulate a general formalism for the calculation of persistence ranges in Sec.~\ref{genMethod}.
In Secs.~\ref{genOmnivore} and \ref{Generalists} we use this formalism to investigate the dependence of the persistence range on the predator-prey interactions and on the geographical network.
We conclude with a summary and discussions in Sec.~\ref{Sum}.

\section{Modeling framework}
\label{MetaCommTheory}
Following Pillai et al.\ \cite{pillai2010,pillai2011} we consider a metacommunity consisting of $s$ species that populate a network of discrete, interconnected patches.
We do not account for the abundance in specific patches, but capture the presence (or absence) of a certain species in a particular patch by a binary variable. Thus every individual patch is either empty or occupied by a specific set of species. 

The system evolves due to local extinctions and colonization of adjacent patches. 
Local extinctions occur either spontaneously (e.g.~due to external events or demographic stochasticity) or due to interactions with other species in the model.  
For every species $i$ we denote the rate of spontaneous local extinctions as $e_i$. 
When the extinction of a species occurs, all species directly or indirectly feeding upon this species also go locally extinct. For example, in a patch with three species where species 3 feeds on species 2 and species 2 feeds on species 1, species 1 goes extinct with rate $e_1$, species 2 goes extinct with rate $e_1+e_2$ and species 3 goes extinct with rate $e_1+e_2+e_3$.

Colonization allows a species $i$ that is established in a patch to establish itself in suitable patches with a constant rate $c_i$. 
A target patch is considered suitable if a) it can be reached from the source patch, b) no stronger competitor is already established in the target patch, and c) prey for the focal species is established in the target patch.

Regarding a) we assume that the patches form a complex network, to which we refer as the geographical network. In this network every node represents a patch, and a link between two nodes indicates that colonization is possible between the corresponding patches. We assume that the geographical network is identical for all species, which is not necessarily true in all systems. Furthermore, we note that assuming constant colonization rates per link implies that dispersal is not strongly limited by the number of colonizers.   

Regarding competition, b) we assume that specialists are better adapted and therefore superior competitors than generalists. If there is an exploitative competition between a specialist and a generalist predator in the same patch and the generalist cannot feed on the specialist then the generalist is excluded. This is assumed to occur instantly in the model. Thus, a generalist predator cannot colonize a patch where a competing specialist is already established. Furthermore, an established generalist population goes locally extinct if a competing specialist colonizes its patch. 

Regarding the availability of suitable prey, c), we assume that only certain primary species can colonize empty patches.
All other species require suitable prey species in their respective patches. In any particular model system we thus choose a fixed set of suitable prey species for each species in the model.  Together all of these potential predator-prey relationships also define a network. In this trophic network the nodes represent the species, whereas directed links represent potential predator-prey interactions. We note that every food chain observed in a particular patch is a subgraph of this trophic network (i.e.\ a subset of its nodes and links). Therefore, also the food web that is formed on the landscape scale, by superposition of the food chains found in patches, is a subgraph of the trophic network.   
The trophic network is thus the maximal food web that can be observed on the landscape scale. However, the observed food webs may be missing some nodes, if some species go extinct on all patches, or links, if some predator-prey interactions are not realized in any patch.   

In the following sections we investigate the conditions for the coexistence of species on the landscape scale, depending on the structure of the trophic and geographical networks. 

\section{Linear food chain}
\label{ChainTop}
We start by considering the case where the maximal trophic network is a linear tri-trophic food chain (see Fig.~\ref{singlebr} a)).
The main purpose of this simple example is to illustrate the calculation of coexistence ranges, which is extended into a general rule for more complex cases in the subsequent sections.  
We note that a very similar system, although on a different network, was already analyzed in \cite{pillai2011}.

To gain a mathematical understanding of the metacommunity dynamics we formulate a mean-field model describing the density of patches in which a specific set of species is found. 
We use a notation where symbols of the form $[i]$ indicate the proportion of patches where the local food chain has length $i$. Thus, $[0]$, $[1]$, $[2]$ and $[3]$ denote the proportion of patches that are empty, inhabited only by species 1, inhabited only by species 1 and species 2, and inhabited by all three species, respectively.
Conservation of the number of patches implies $[0]+[1]+[2]+[3]=1$.
Additionally we use $p_1$, $p_2$ and $p_3$ to denote the proportion of patches where species 1, species 2 and species 3 is present, irrespective of the presence of other species. Consequently, $p_1=[1]+[2]+[3], p_2=[2]+[3]$ and $p_3=[3]$.

The expected time evolution of the mean-field patch densities is then given by
\begin{linenomath}
\begin{align}
\label{chaineq}
\begin{split}
 \tfrac{d}{dt}[0]&=-c_1\langle k \rangle[0]p_1+e_1p_1\\
 \tfrac{d}{dt}[1]&=c_1\langle k \rangle[0]p_1-c_2 \langle k \rangle [1]p_2+e_2p_2-e_1[1]\\
  \tfrac{d}{dt}[2]&=c_2\langle k \rangle[1]p_2-c_3 \langle k \rangle [2]p_3+e_3p_3-(e_1+e_2)[2]\\
  \tfrac{d}{dt}[3]&=c_3\langle k \rangle[2]p_3-(e_1+e_2+e_3)[3],
  \end{split}
\end{align}
\end{linenomath}
where we introduced $\langle k \rangle$ to denote the mean number of neighboring patches per patch.

In Eq.~(\ref{chaineq}) the first terms on the right-hand-sides describe the effect of colonization, whereas the second terms describe the effect of extinctions. 
To understand the specific functional form appearing in the equations let us consider the first term of the first equation, $-c_1\langle k \rangle[0]p_1$, which describes the loss of empty patches due to colonization. We lose empty patches due to colonization from adjacent patches in which species 1 is established. A typical patch has $\langle k \rangle$ adjacent patches and species 1 is established in any of these with probability $p_1$. 
Multiplying the expected number of adjacent patches where species 1 is established, $p_1\langle k \rangle$, with the 
colonization rate $c_1$ yields the total expected colonization rate for a given empty patch. Multiplying further with the proportion of empty patches, $[0]$, yields the total loss rate $c_1\langle k \rangle[0]p_1$.   

We note that in the equations for empty patches ($[0]$) and tri-trophic patches ($[3]$) the effect of colonization is purely negative and purely positive, respectively.
By contrast, for single-species $([1])$ and two-species $([2])$ patches colonization appears both, as a gain and as a loss term, as patches of these types can be both, created and destroyed by colonization.  
Similarly, the effect of extinctions is purely positive for $[0]$-patches and purely negative for $[3]$-patches, whereas for $[1]$- and $[2]$-patches extinction has both, positive and negative effects.  

In the equations we replaced the number of patches adjacent to a given patch, with the expectation value of adjacent patches $\langle k \rangle$, the so-called \emph{mean degree}. Working with this mean degree instead of the full probability distribution of the numbers of adjacent patches (the \emph{degree distribution}) is known as \emph{homogeneous approximation} in network science. This approximation is known to yield good results for all reasonably narrow degree distributions, which should cover most ecologically relevant cases. However, we emphasize that a somewhat more complicated computation is necessary for instance for geographical networks with a central hub.  

By setting the right-hand sides of Eq.\,\eqref{chaineq} to zero and solving the resulting system of equations, we find the expected equilibrium patch densities 
\begin{linenomath} 
\begin{align*}
[0]^*&= \frac{e_1}{c_1\langle k \rangle}\\
[1]^*&= \frac{e_1+e_2}{c_2\langle k \rangle}\\
[2]^*&= \frac{e_1+e_2+e_3}{c_3\langle k \rangle}\\
[3]^*&=1-\frac{e_1}{c_1\langle k \rangle}- \frac{e_1+e_2}{c_2\langle k \rangle}-\frac{e_1+e_2+e_3}{c_3\langle k \rangle}=1-\sum_{i=0}^{2} [i]^*,
\end{align*}
\end{linenomath}
where we used asterisks to indicate the stationary values that the system approaches after sufficiently long time.

Generalizing these results to a class of systems where the maximal network is a linear food chain of length $l$, one finds
\begin{linenomath}
\begin{align*}
[i]^*&= \frac{e_1+\dots+e_{i+1}}{c_{i+1}\langle k \rangle} ,  \qquad\quad 0\le  i < l\\
[l]^*&=1-\sum_{i=0}^{l-1} [i]^*.
\end{align*}
\end{linenomath}

In the following we use similar reasoning to study more complex cases. 
To simplify the presentation we assume that the extinction rates for all species are equal, $e_i=e\; \forall i$, and further that the colonization rates for all species are equal, $c_i=c\; \forall i$. 
Additionally, we introduce the dimensionless parameters $z=e/c$ and $\bar z=z/\langle k \rangle$, which constitute differently normalized extinction rates. 
In this notation the equilibrium patch densities are
\begin{linenomath}
\begin{align}
[i]^*&= (i+1)\bar z,  && 0\le  i < l\nonumber\\
\label{eqdens}
[l]^*&=1-\tfrac{1}{2}l(l+1)\bar z, &&\\
p_i^*&=1-\tfrac{1}{2}i(i+1)\bar z, && 1\le  i \le l.
\nonumber
\end{align}
\end{linenomath}

Let us now investigate under which conditions a food chain of length $l$ can persist in the metacommunity.
To obtain an answer directly, we use an approach motivated by the epidemic literature \cite{kamgang2008,vandriessche2002,diekmann2010,boehme2011}
and compute the linear stability of the ``$l$-free equilibrium", i.e.~the equilibrium where the species $l$ is absent, which is analogous to considering the ''disease-free equilibrium`` for epidemics. 
In other words, we analyze the stability of a food chain of length $l-1$ against arrival of species $l$. 
This reveals the threshold parameter values at which the $l-1$-trophic food chain becomes unstable and species $l$ can establish itself. 
As will become apparent below, this is also the threshold for persistence of species $l$ because the present model cannot exhibit Allee-effects on the metapopulation level.

From the evolution equation for the patch density $[l]$,
\begin{linenomath}
\begin{equation*}
\tfrac{d}{dt}[l]=(c\langle k \rangle [l-1]-le)[l],
\end{equation*}
\end{linenomath}
it follows that persistence of species $l$ is possible if
\begin{linenomath}
\begin{equation}
\label{stabcond}
(c\langle k \rangle [l-1]^*-le)[l]>0.
\end{equation}
\end{linenomath}
We note that $[l-1]^*$ denotes the equilibrium patch density of species $l-1$ in the ``$l$-free equilibrium'', before arrival of species $l$.
The condition Eq.~\eqref{stabcond} shows that the prey, species $l-1$, must reach a threshold density $[l-1]^* = l\bar{z}$ to allow for persistence of species $l$, which confirms basic ecological intuition.
 
Using Eq.\,\eqref{eqdens} the persistence condition for $l$ can be rewritten as
\begin{linenomath}
\begin{equation}
\label{specinv}
z < \frac{2\langle k \rangle}{l(l+1)},
\end{equation}
\end{linenomath}
which can be read as a bound for the dimensionless extinction rate $z$.
Thus, high extinction rates hinder the formation of long food chains while a dense geographical connectivity (high $\langle k \rangle$) promotes it. 

\section{Omnivory in a tri-trophic example}
\label{Omnivores}
In this section, we calculate the parameter range for $z$ where food webs containing specialists and omnivores can persist at the metacommunity scale. These analyses generalize the results of \cite{pillai2011} by providing a general formalism for the computation of persistence ranges.

\begin{figure}
\begin{center}
\includegraphics[trim=4.5cm 18cm 6.5cm 6cm,width=.8\textwidth]{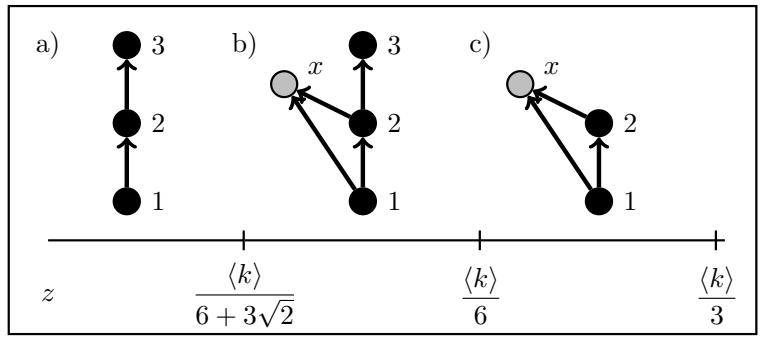}
\caption{Example for a maximal food web emerging in the metacommunity. Specialist species are depicted by black nodes, omnivores by gray nodes. Specialists are numbered according to the trophic level they belong to. Arrows indicate feeding links. 
Shown are different food web that arise on the landscape-scale in the parameter rangesindicated. c) The omnivore $x$ is able to persist in a system of two specialists for $z<\langle k \rangle/3$, coexists with the specialist species 3
for $z<\langle k \rangle/6$ and becomes extinct due to competitive exclusion for $z<\langle k \rangle/(6+3\sqrt{2})$. These parameter ranges are not drawn to scale. See text for derivation.}
\label{singlebr}
\end{center}
\end{figure}

As a first example, we consider the food web configuration shown in Fig.~\ref{singlebr}. Here, the maximal food web consists of a linear food chain and an additional omnivore top predator that can feed on both the producer and the primary consumer in the food chain.
We investigate under which conditions the omnivore, $x$, can coexist with the specialist top-predator of the food chain, although the specialist will exclude the omnivore in any given patch. 

Even without mathematical reasoning it is intuitive that the omnivore faces two dangers. First, if the geographical network is very sparse then the prey species may be too sparse to allow for persistence of the omnivore. Second, if the geographical network is too dense then the specialist top-predator will establish, which threatens the omnivore with competitive exclusion.

In the following we mathematically determine the threshold connectivities of the geographical network by employing the same line of reasoning as for the linear chain. 
We start by writing the governing equations for patches in which the omnivore $x$ is established as 
\begin{linenomath}
\begin{align}
\label{omniv12}
\begin{split}
\tfrac{d}{dt}[1x]&=c\langle k \rangle([1]p_x-[1x]p_2)+e([2x]-2[1x])\\
\tfrac{d}{dt}[2x]&=c \langle k \rangle ([2]p_x+[1x]p_2-[2x]p_3)-3e[2x],
\end{split}
\end{align}
\end{linenomath}
where we used symbols of the form $[ix]$ to denote patches in which a specialist food chain of length $i$ and the omnivore $x$ is established.
Note that omnivore $x$ does not suffer competitive exclusion in $[1x]$-patches, because species 2 is not only a competitor for $x$, but also one of its prey species.
Thus, colonization of species 2 turns a $[1x]$-patch into a $[2x]$-patch. This contribution is captured in the term $c \langle k \rangle [1x]p_2$.

To allow for a more concise treatment we can rewrite Eqs.~\eqref{omniv12} in the matrix form
\begin{linenomath}
\begin{equation}
\label{matrixeq}
\frac{d}{dt}
 \begin{pmatrix}
[1x]\\\\
[2x]
\end{pmatrix}=
\underbrace{
 \begin{pmatrix}
 c\langle k \rangle ([1]-p_2)-2e & c \langle k \rangle[1]+e\\&\\
 c\langle k \rangle ([2]+p_2) &  c\langle k \rangle ([2]-p_3)-3 e
 \end{pmatrix}
}_{\mathbf{S}}
 \begin{pmatrix}
[1x]\\\\
[2x]
\end{pmatrix},
\end{equation}
\end{linenomath}
describing the time evolution of the vector $\vec{p}_x=([1x],[2x])$. 

Now, we analyze the response of the $x$-free equilibrium to small perturbations, i.e.\ arrival of $x$ in a small number of patches.
Persistence of the omnivore $x$ is possible when the $x$-free equilibrium is unstable. 
From nonlinear dynamics we know that this is the case when the largest eigenvalue $\lambda$ of the matrix $\mathbf{S}$, in the $x$-free equilibrium, is positive, 
which we write as 
\begin{linenomath}
\begin{equation}
\label{smatrix-omniv12}
\lambda(\mathbf{S}^*)>0.
\end{equation}
\end{linenomath}
Starting from this condition we now extract an explicit condition in terms of the connectivity of the geographical network $\langle k \rangle$ and the extinction rate $z$.

In the evaluation of the eigenvalue inequality, we have to take into account that $\mathbf{S}^*$ depends on the equilibrium patch density of species 3. Because it thus makes a difference whether species 3 is present or absent, we have to analyze the stability of the $x$-free equilibrium separately for these two cases (see Fig.~\ref{singlebr} b) and c)). To distinguish between the two cases, we use $\mathbf{\check{S}^*}$ to denote $\mathbf{S}^*$ when species 3 is present ($0\le z < \langle k \rangle/6$) and $\mathbf{\hat{S}^*}$ to denote $\mathbf{S}^*$ when species 3 is absent ($\langle k \rangle/6 \le z \le \langle k \rangle/3$). 
The persistence range for the omnivore $x$ is then obtained from a combination of the two conditions $\lambda(\mathbf{\check{S}}^*)>0$ and $\lambda(\mathbf{\hat{S}}^*)>0$ .

Let us first consider the case when species 3 is present. Inserting the corresponding equilibrium expressions from Eqs.~\eqref{eqdens} into Eq.~\eqref{matrixeq}, we can write 
\begin{linenomath}
\begin{align}
\label{smat1}
\mathbf{\check{S}}^*&=
\underbrace{
\begin{pmatrix}
 -1 &0\\1 & -1
\end{pmatrix}
}_{\mathbf{\check{s}_1}}\langle k \rangle 
+
\underbrace{
\begin{pmatrix}
 3 &3\\0 & 6
\end{pmatrix}
}_{\mathbf{\check{s}_2}}z.
\end{align}
\end{linenomath}
The matrix $\mathbf{\check{S}}^*$ is thus split into two matrices, $\mathbf{\check{s}_1}$ and $\mathbf{\check{s}_2}$, which capture the net balance of $x$-patches due to colonization and local extinction, respectively. 

From \eqref{smat1} it is clear that the x-free equilibrium is stable at $z=0$, because all eigenvalues of $\mathbf{\check{S}^*}_{z=0}=\mathbf{\check{s}_1}$ are negative. 
When $z$ is increased a threshold $z^\prime$ is eventually reached where one eigenvalue becomes positive and the omnivore can no longer persist.
Mathematically we can detect the transition by monitoring the determinant $\det(\mathbf{\check{S}^*})$, which becomes negative at $z^\prime$. In order to calculate the threshold, we write
\begin{linenomath}
 \begin{align}
 \label{dets}
 \det(\mathbf{\check{S}^*})=\det(\mathbf{\check{s}_1}\langle k \rangle+\mathbf{\check{s}_2}z)=\det(-\mathbf{\check{s}_1}) \det(\mathbf{\check{M}}z-\langle k \rangle),
  \end{align}
\end{linenomath}
  where $\mathbf{\check{M}}=-\mathbf{\check{s}_2}\mathbf{\check{s}_1}^{-1}$, and $\mathbf{\check{s}_1}^{-1}$ is the inverse matrix of $\mathbf{\check{s}_1}$.
  From Eq.~\eqref{dets} it follows that
 $\det(\mathbf{\check{S}^*})<0$ if either a) $\det(-\mathbf{\check{s}_1})<0$ and all (real) eigenvalues of $\mathbf{\check{M}}$ smaller than $\langle k \rangle/z$, or
 b) $\det(-\mathbf{\check{s}_1})>0$ and $\lambda(\mathbf{\check{M}})>\langle k \rangle/z$, where $\lambda(\mathbf{\check{M}})$ is the largest (real) eigenvalue of $\mathbf{\check{M}}$. As for the matrix given in Eq.~\eqref{smat1} we have $\det(-\mathbf{\check{s}_1})>0$, the omnivore persists in the system for $z>z^\prime=\langle k \rangle/\lambda(\mathbf{\check{M}})$.
  
Now we have an explicit expression for the threshold $z^\prime$. For the matrix $\mathbf{\check{S}}^*$ in Eq.\,\eqref{smat1},
\begin{linenomath}
\begin{equation*}
\mathbf{\check{M}}=
\underbrace{
\begin{pmatrix}
 3 &3\\0 & 6
\end{pmatrix}
}_{\mathbf{\check{s}_2}}
\cdot
{\underbrace{
\begin{pmatrix}
 1 &0\\1 & 1
\end{pmatrix}}_{-\mathbf{\check{s}_1}^{-1}}}
=\begin{pmatrix}6&3\\6&6\end{pmatrix},
\end{equation*}
and therefore, the persistence condition for $p_3\neq0$ becomes
\begin{equation}
\label{persx}
z>\frac{\langle k \rangle }{6+3\sqrt{2}}.
\end{equation}
\end{linenomath}
This is a lower bound for the parameter $z$, which limits the persistence of $x$ due to competition with the specialist 3.

Now, we consider the case $p_3^*=0$, shown in Fig.~\ref{singlebr} c). Inserting the corresponding expressions from Eqs.~\eqref{eqdens} in Eq.\,\eqref{matrixeq} yields
\begin{linenomath}
\begin{equation}
\label{smat2}
\mathbf{\hat{S}}^*=
\underbrace{
\begin{pmatrix}
 -1 &0\\2 & 1
\end{pmatrix}
}_{\mathbf{\hat{s}_1}}\langle k \rangle
+
\underbrace{
\begin{pmatrix}
 3 &3\\-6 & -6
\end{pmatrix}
}_{\mathbf{\hat{s}_2}} z.
\end{equation}
\end{linenomath}
In this case, we observe that the stability of the $x$-free equilibrium is lost when approaching some threshold value $z^\prime$ from above. This can be understood intuitively, because the $x$-free equilibrium is certainly stable when there is no prey species 2, i.e.\ for $z\ge\langle k \rangle/3$. It becomes unstable if one eigenvalue of $\mathbf{\hat{S}^*}$ becomes positive at a value $z^\prime$, and accordingly $\det(\mathbf{\hat{S}^*})$ becomes negative. 
An analogous argumentation as before in Eq.~\eqref{dets} leads to the conclusion that $\lambda(\mathbf{\hat{S}^*})>0$ is fulfilled when $z<z^\prime=\langle k\rangle/\lambda(\mathbf{\hat{M}})$, where $ \mathbf{\hat{M}}=-\mathbf{\hat{s}_2}\mathbf{\hat{s}_1}^{-1}$ and we used that  $\det(-\mathbf{\check{s}_1})<0$ for the matrix given in Eq.~\eqref{smat2}.

For the matrix $\mathbf{\hat{S}}^*$ in \eqref{smat2} the corresponding condition yields
\begin{linenomath}
\begin{equation}
\label{invx}
z<\frac{\langle k \rangle}{3}.
\end{equation}
\end{linenomath}
This constitutes an upper bound for $z$, below which omnivore $x$ persists in a system of two specialists.
Note that we obtain the same upper bound for the persistence range of a specialist species 2, according to Eq.\,\eqref{specinv}. 
Thus, as soon as a chain of two specialists exists in the metacommunity, an omnivore feeding upon both of them can establish.
Below we show that this holds for any omnivore feeding upon two prey species on subsequent trophic levels. 

Taking conditions Eq.\,\eqref{persx} and Eq.\,\eqref{invx} together, we conclude that for
\begin{linenomath}
\begin{equation}
\label{coexx}
\frac{\langle k \rangle }{6+3\sqrt{2}}<z<\frac{\langle k \rangle}{3}
\end{equation}
\end{linenomath}
omnivore $x$ is able to coexist with a chain of specialists at the metacommunity level. 

A summary of the results for this example system are shown in Fig.~\ref{singlebr}. We found that for $z<\langle k \rangle /3$ omnivore $x$ can establish itself in the metacommunity.
In contrast to the persistence range of species 2, the persistence range of omnivore $x$ is bounded from below, due to competition with specialist 3, which can persist in the system for $z<\langle k \rangle /6$. Omnivore $x$ can coexist with the superior competitor 3 for $z>\langle k \rangle/(6+3\sqrt{2})$ (b), while
for $z<\langle k \rangle/(6+3\sqrt{2})$ the omnivore becomes extinct and the observed food web is a linear chain of three specialists (a).

The results obtained in this section reproduce those already obtained by \cite{pillai2011}.
Whereas \cite{pillai2011} obtained these results from the solution of the full underlying equation system, we compute them directly from a stability analysis. The advantage of this approach is that it can be easily applied to more complicated situations and lends itself to automation in computer algebra systems, which allows to compute coexistence ranges for a large variety of different food webs.  

\begin{figure}
\framebox{\parbox{\textwidth}{
\begin{enumerate}
\item Construct matrix $\mathbf{S}$ from evolution equations for $\vec{p_x}$.
\item Insert the corresponding equilibrium densities from Eqs.~\eqref{eqdens} in order to obtain $\mathbf{\check{S}}^*$ and $\mathbf{\hat{S}}^*$.
\item Write $\mathbf{\check{S}}^*$ in the form $\mathbf{\check{s}_1}\langle k \rangle + \mathbf{\check{s}_2} z$. Do the same for $\mathbf{\hat{S}}^*$.
\item Determine $\mathbf{\check{M}}= -\mathbf{\check{s}_2}\mathbf{\check{s}_1}^{-1}$ and $\lambda(\mathbf{\check{M}})$. Do the same for $\mathbf{\hat{M}}$.
\item The persistence condition is $\langle k \rangle /\lambda(\mathbf{\check{M}})<z<\langle k \rangle /\lambda(\mathbf{\hat{M}})$. 
\end{enumerate}
}}
\caption{Recipe for the calculation of persistence thresholds.}
\label{recipebox}
\end{figure}
\section{General method for persistence thresholds}
\label{genMethod}
We now exploit the power of the method proposed in the previous section to obtain general results on the persistence of omnivores. Consider an omnivore feeding on different prey species, where the trophic level of the highest prey species is $i-1$. In this case the corresponding $\mathbf{S^*}$-matrix is given by
\begin{linenomath}
\begin{equation*}
 \mathbf{S^*}=
 \begin{cases}
             \mathbf{\check{S}^*}=\mathbf{\check{s}_1}\langle k \rangle +\mathbf{\check{s}_2}z, \,\,
              \quad \textnormal{for} &0\le z < \displaystyle\frac{ 2\langle k \rangle}{i(i+1)}, \\
&\\
 \mathbf{\hat{S}^*}=\mathbf{\hat{s}_1}\langle k \rangle +\mathbf{\hat{s}_2}z,  \,\,
	\quad \textnormal{for}  &\displaystyle\frac{2\langle k \rangle}{i(i+1)} \le z \le \displaystyle\frac{2 \langle k \rangle}{i(i-1)}.\\
            \end{cases}
\end{equation*}
\end{linenomath}
The different ranges appear in this equation because of the extinction of species $i$ and extinction of species $i-1$, respectively.
From the general expression for $\mathbf{S^*}$ we can proceed analogously to the previously discussed specific example.
Importantly, certain properties of the specific example system that we exploited above remain valid in the general case.
For example, $\mathbf{\check{S}^*}$ has only negative eigenvalues at $z=0$, because $\mathbf{\check{s}_1}$ is always lower triangular and has negative diagonal entries. Therefore $\det(-\mathbf{\check{s}_1})>0$, and due to Eq.~\eqref{dets} there is a lower bound for the persistence range, which is determined by $\lambda(\mathbf{\check{M}})$. Conversely, as $\det(-\mathbf{\hat{s}_1})<0$, there is an upper bound of the persistence range, if $\lambda(\mathbf{\hat{M}})=\langle k \rangle/z$ in the relevant range. Otherwise, the persistence range is naturally bounded at $2 \langle k \rangle/(i(i-1))$, where the highest prey species of the omnivore (species $i-1$) goes extinct.  
 
Summarizing the considerations above, we find
\begin{linenomath}
\begin{equation}
\label{prop}
 \lambda(\mathbf{S^*})>0\, \quad\textnormal{  in the range }\quad
\frac{\langle k \rangle}{\lambda(\mathbf{\check{M}})}<z<\frac{\langle k \rangle}{\lambda(\mathbf{\hat{M}})}.
\end{equation}
\end{linenomath}
This relationship enables the direct computation of the persistence range for a given omnivore, and thus allows for the analysis of a broad class of food webs in metacommunities. An overview of this procedure is given in Fig.~\ref{recipebox}. An additional example for the application of this procedure to a specific food web is given in the Appendix. 

\section{General results on omnivory}
\label{genOmnivore}
Let us now use the method proposed above to gain ecological insights into the persistence of omnivores in food chains.
For simplicity, we restrict our analysis to omnivores feeding upon two prey species, but note that the proposed method can be likewise applied to omnivores feeding on more prey species.
Unless stated otherwise we denote the trophic levels of the omnivores' prey by $i-1$ and $j-1$, where $i>j$. 
Correspondingly, the specialists competing with the respective omnivore occupy trophic levels $i$ and $j$.

\begin{figure}
\begin{center}
\includegraphics[width=.8\textwidth,trim=4cm 17cm 6cm 4cm]{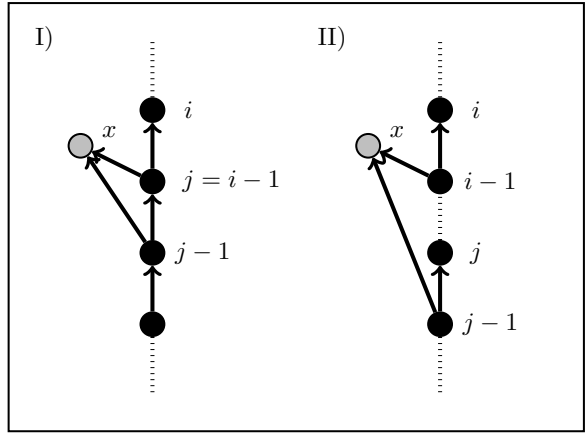}
\caption{General food web configuration consisting of a chain of specialists and an omnivore $x$. In I) the omnivore $x$ feeds on two subsequent trophic levels, i.e.\,$i=j+1$, whereas in II)
the omnivore $x$ feeds on two non-subsequent trophic levels, i.e.\,$i>j+1$. Accordingly, different coexistence ranges are obtained for both cases, as described in the text. }
\label{generalcase}
\end{center}
\end{figure}

We distinguish between the following two cases (see Fig. \ref{generalcase}): 
I) The omnivore feeds upon prey at subsequent trophic levels ($i=j+1$).
II) The omnivore feeds upon prey at non-subsequent trophic levels ($i>j+1$).
The $\mathbf{S}^*$-matrices governing the evolution equations for $[(i-1)x]$- and $[(j-1)x]$-patches in the $x$-free equilibrium are then given by
\begin{linenomath}
\begin{align}
\label{smatrix1}
\mathbf{S}_{\rm I}^*&=
\begin{pmatrix}
 \langle k \rangle ([j-1]^*-p_j^*)-j z & \langle k \rangle[j-1]^*+ z\\
&\\
 \langle k \rangle([i-1]^*+p_{i-1}^*) & \langle k \rangle([i-1]^*-p_i^*)-i z
\end{pmatrix}
\end{align}
\end{linenomath}
and
\begin{linenomath}
\begin{align}
\label{smatrix2}
\mathbf{S}_{\rm II}^*&=
\begin{pmatrix}
 \langle k \rangle([j-1]^*-p_j^*)-j z & \langle k \rangle[j-1]^*+ z\\
&\\
 \langle k \rangle[i-1]^* & \langle k \rangle([i-1]^*-p_i^*)-i z
\end{pmatrix},
\end{align}
\end{linenomath}
where we introduced the subscripts I and II in order to distinguish the $\mathbf{S}^*$-matrices in the different cases.

In order to calculate the coexistence range for a general omnivore, we have to separately consider the situation where species $i$ is absent ($p_i^*=0, [i-1]^*=p_{i-1}^*=1-(i(i-1)\bar z)/2$),
and the situation where species $i$ is present in the system ($p_i^*=1-(i(i+1)\bar z)/2, [i-1]^*=iz$). Then, combining both conditions, we expect to find a parameter range for $z$ which is bounded from below and from above.

Following the procedure in  Fig.~\ref{recipebox} and using Eq.\,\eqref{prop}, we find 
\begin{linenomath}
\begin{align*}
\mathbf{\check M}_{\rm I}&=
 \begin{pmatrix}
  \frac{i(i+1)}{2} & i\\
2i & \frac{i(i+1)}{2}
 \end{pmatrix},&
 \mathbf{\hat M}_{\rm I}&=
  \begin{pmatrix}
   \frac{i(i-5)}{2} & -i\\
 2i & \frac{i(i+1)}{2}
  \end{pmatrix},\\
\mathbf{\check M}_{\rm II}&=
 \begin{pmatrix}
  \frac{j(j+1)}{2} & j+1\\
i & \frac{i(i+1)}{2}
 \end{pmatrix},&
\mathbf{\hat M}_{\rm II}&=
 \begin{pmatrix}
  \frac{(j-2)(j+1)}{2} & -(j+1)\\
i & \frac{i(i+1)}{2}
 \end{pmatrix},
 \end{align*}
\end{linenomath}
and the corresponding coexistence ranges
\begin{linenomath}
\begin{align}
  \label{gencoex1}
\rm I)&&
\frac{\langle k \rangle}{\lambda(\mathbf{\check M}_{\rm I})}<z&<\frac{\langle k \rangle}{\lambda(\mathbf{\hat M}_{\rm I})},&\\
  \label{gencoex2}
\rm II)&&
\frac{\langle k \rangle}{\lambda(\mathbf{\check M}_{\rm II})}<z&<\frac{\langle k \rangle}{\lambda(\mathbf{\hat M}_{\rm II})}.&
\end{align}
\end{linenomath}

\begin{figure}
\begin{center}
\includegraphics[width=.8\textwidth]{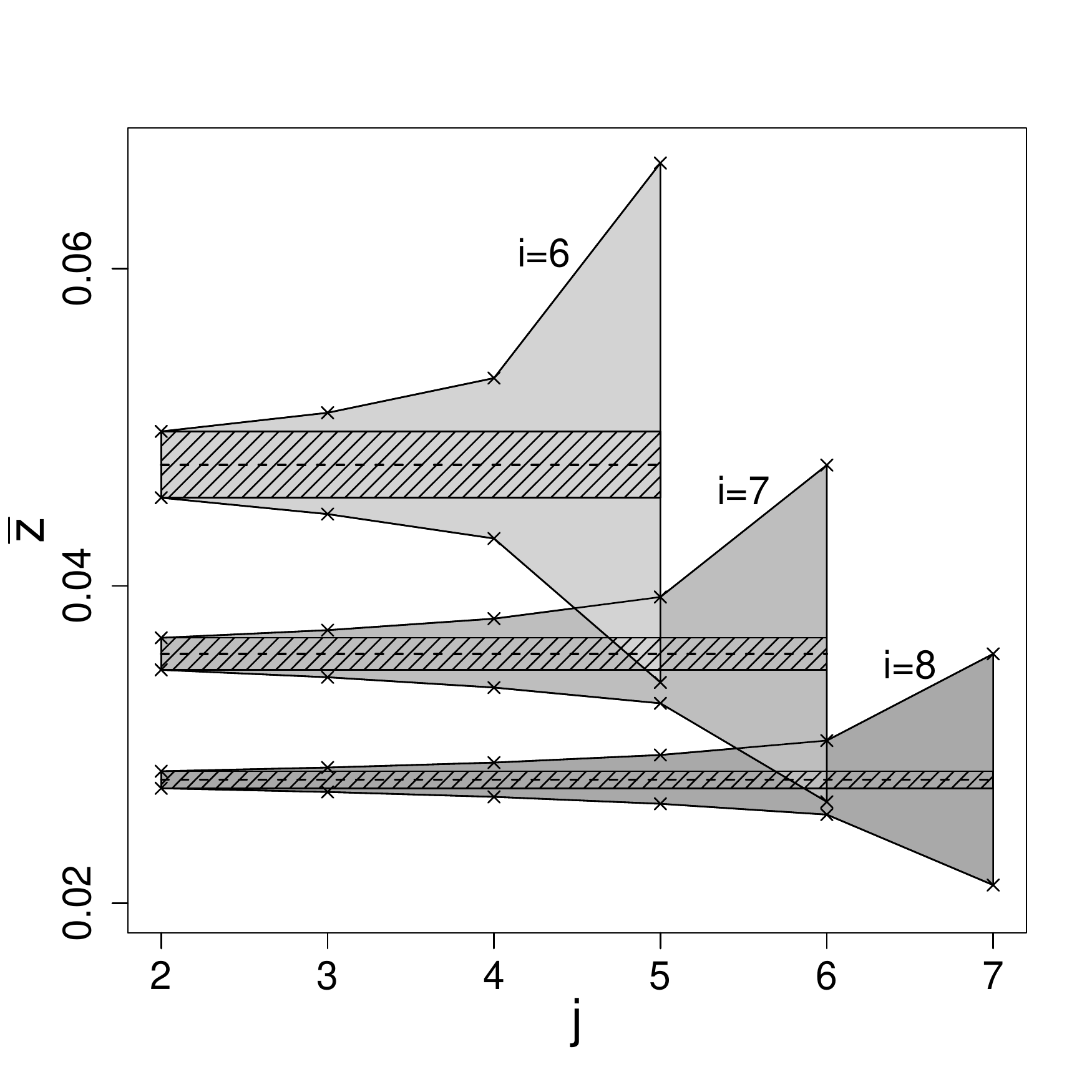}
\caption{Dependence of the coexistence range for an omnivore on the trophic levels of its prey. The grey regions show the coexistence ranges for different values of $i$ and varying $j$, corresponding to an omnivore feeding on species $i-1$ and $j-1$ (see Eqs.\,\eqref{gencoex1},\eqref{gencoex2}). The coexistence range becomes smaller with increasing $i$ and for fixed $i$ with decreasing $j$. This shows that omnivores are more likely to coexist if they feed on adjacent tropic levels, low in the food chain.   
Dashed lines indicate the respective persistence thresholds of species $i$. 
The dashed region marks the parameter range where the maximal number of omnivores coexists with the specialist chain.}
\label{ijdep}
\end{center}
\end{figure}

Equations \eqref{gencoex1} and \eqref{gencoex2} reveal the impact of the connectivity of the patch-network, showing that dense topologies promote the persistence of omnivores in the metacommunity.
We can further deduce statements concerning the dependence of the coexistence range on the trophic levels of the prey species:
\begin{enumerate}
\item The coexistence range for omnivore $x$ becomes smaller with increasing trophic level of the prey species.
\item The coexistence range for omnivore $x$ becomes smaller with increasing distance between the trophic levels of its prey species.
\end{enumerate}
These characteristics of the coexistence range are shown in Fig.~\ref{ijdep}.
Note that if we are only interested in the range where the omnivore coexists with the competitor $i$,
the upper thresholds in Eq.\,\eqref{gencoex1} and Eq.\,\eqref{gencoex2} become the persistence threshold for species $i$, i.e.\,$z<2 \langle k \rangle/(i(i+1))$ (dashed line in Fig.~\ref{ijdep}). The range where possibly more than one omnivore coexists with a chain of specialists is given by the intersection of the respective individual coexistence ranges. The dashed areas in Fig.~\ref{ijdep} mark the parameter ranges where all coexistence ranges for the same value of $i$ overlap. Thus the maximal number of coexisting omnivores in these parameter ranges yields $n_x^{max}=i-1$.

Let us emphasize that the coexistence range of an omnivore only depends on the trophic levels of its prey species and not on the total length of the specialist chain.
This is a consequence of the following relationship:
\begin{linenomath}
 \begin{align}
 \label{rel}
  [i-1]+[(i-1)x]=
 \begin{cases}
  i\bar z;\qquad \qquad \qquad \quad \, p_i\neq 0\\
 1-\tfrac{1}{2}i(i-1)\bar z;\qquad p_i=0.
 \end{cases}
 \end{align}
\end{linenomath}
In the $x$-free equilibrium, i.e.\,for $[(j-1)x]=[(i-1)x]=0$, the equilibrium densities given in Eq.\,\eqref{eqdens} are recovered.
Relation \eqref{rel} can be interpreted as ``biomass-conservation'', implying that the species densities $p_{i-1}$ in the chain are conserved, no matter how many predators feed upon species $i-1$ (and upon prey at trophic levels below $i-1$). 
 
\begin{figure}
\begin{center}
\includegraphics[width=.8\textwidth]{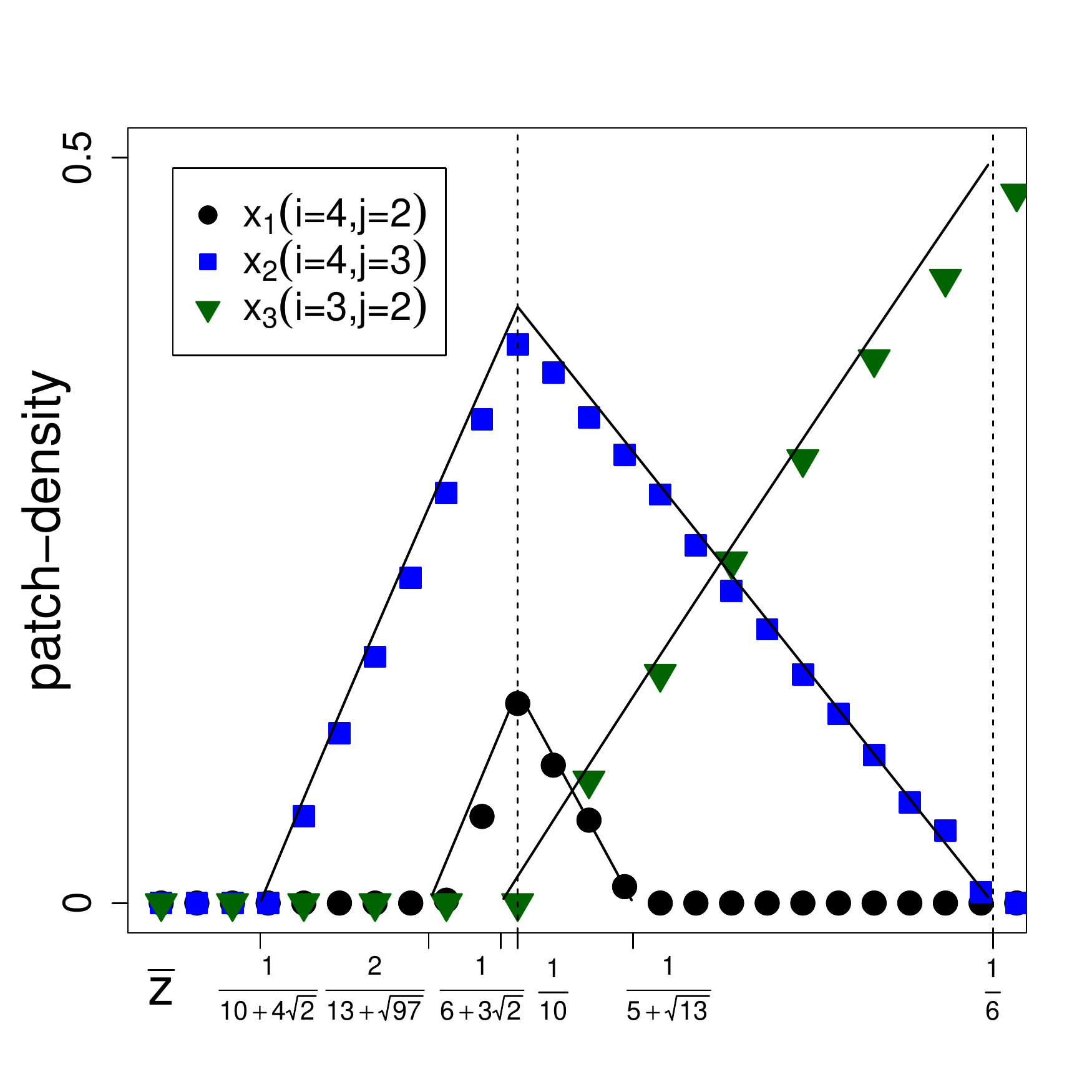}
\caption{Equilibrium densities of three different omnivores from numerical simulations (symbols) and theoretical predictions (lines). Dashed lines indicate parameter values where the respective specialist competitor goes extinct in the metacommunity. The equilibrium patch densities observed in simulations display a close to linear behavior and are well described by the analytical formula in Eq.\,\eqref{pxeq}. The parameters for the simulation are the same as in Fig.~\ref{sim_coex}.}
\label{pxcompare}
\end{center}
\end{figure}

The conservation law in Eq.\,\eqref{rel} has another practical implication: it allows for the direct computation of equilibrium patch densities $p_x$ for omnivores.
In the Appendix we derive that the total density of patches where omnivore $x$ is established is given by 
\begin{linenomath}
  \begin{equation}
 \label{pxeq}
 p_x^*=[(j-1)x]^*+[(i-1)x]^*=\lambda(\mathbf{S}^*), 
 \end{equation}
\end{linenomath}
where $\mathbf{S}^*$ is the corresponding transition matrix in the $x$-free equilibrium (matrices given in Eq.\,\eqref{smatrix1}, Eq.\,\eqref{smatrix2}).
 Note that $\mathbf{S}^*$ is altered at the point where species $i$ enters the system. Thus, there is no unique matrix $\mathbf{S}^*$ describing $p_x^*$ in the whole coexistence range. This is apparent in Fig.~\ref{pxcompare} at the parameter value $\bar{z}=2/(i(i+1))$. So, from the point where invasion of an omnivore becomes possible, at the upper bound of the coexistence range, its patch density increases up to the point where the specialist competitor becomes able to persist in the system. At this point (dashed lines in Fig.~\ref{pxcompare}), the patch density of the omnivore starts to decrease due to competition with the specialist $i$. Both, increase and decrease proceed almost linearly, as it can be seen in Fig.~\ref{pxcompare}. For $i>j+1$ and $p_i^*\neq0$ the linearity is exact because $\mathbf{\check{S}_{\rm II}}^*=-\mathbf{1}+\mathbf{\check M}_{\rm II}\bar z$.

The analytical expression for the equilibrium patch density of omnivores is useful for the computation of persistence conditions of predators that are capable of feeding on other omnivores. Then, the $\mathbf{S}$-matrix contains patch densities of omnivores, in addition to patch densities of specialists.

\section{Competition between generalists and specialists}
\label{Generalists}
\begin{figure}
\begin{center}
\includegraphics[width=.8\textwidth,trim=4cm 18cm 6cm 4cm]{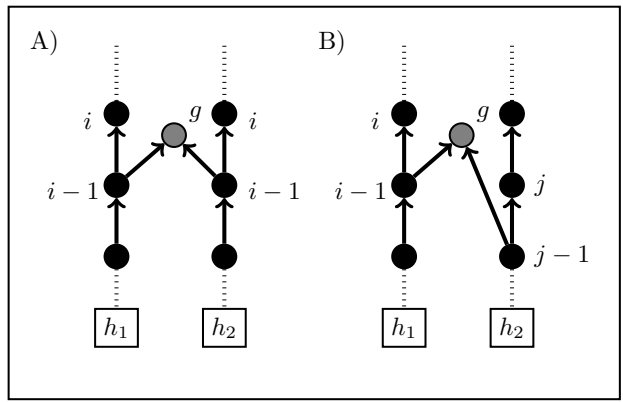}
\caption{General food web configuration consisting of a chain of specialists and a generalist $g$. The generalist feeds on two prey species in two independent chains arising from two types of habitat. 
In A), the trophic level of both prey species is the same ($i=j$), in B) the trophic levels differ at least by one ($i>j$). The corresponding coexistence ranges for both configurations are derived in the text.}
\label{doublebr}
\end{center}
\end{figure}
So far, we considered all empty patches to be equal. In this section we consider a system which comprises two different types of habitat $h_1$ and $h_2$, which we assume to occur with equal probablity.
Primary producers are assumed to specialize on a specific type of habitat. 
Thus, each habitat type sustains a distinct chain of specialists (see Fig.~\ref{doublebr}).
Now, we can ask for the conditions under which a generalist, feeding on prey from both chains, is able to persist in the metacommunity. 

For simplicity, we focus again on generalists feeding on exactly two prey species.
In analogy to the previous analysis we denote the trophic levels of the prey species of the generalist $g$ with $i-1$ and $j-1$, where $j\le i$.
The corresponding matrix $\mathbf{S}^*$, describing the evolution of $[(j-1)g]$- and $[(i-1)g]$-patches in the $g$-free equilibrium is then given by
\begin{linenomath}
\begin{align}
\label{smatrix3}
\mathbf{S}^*&=
\begin{pmatrix}
 \langle k \rangle([j-1]^*-p_j^*)-j z &  \langle k \rangle[j-1]^*\\
&\\
  \langle k \rangle[i-1]^* &  \langle k \rangle([i-1]^*-p_i^*)-iz
\end{pmatrix}.
\end{align}
\end{linenomath}
In contrast to above, $[(j-1)g]$-patches cannot become $[(i-1)g]$-patches or vice versa, as both prey species are from different, independent chains.

For the calculation of the coexistence range, we distinguish the two cases (see Fig. \ref{doublebr}):
A) The two prey species are at the same trophic level.
B) The two prey species are at different trophic levels.
This is necessary because generalists feeding upon two prey species at the same level, say $i-1$, experience no competition with specialists if species $i$ is not present. In contrast, a generalist feeding upon one prey at level $i-1$ and a second one at some lower 
level $j-1$, even in the absence of specialist $i$, experiences competition from the specialist $j$.

Proceeding according to Fig.~\ref{recipebox}, we find 
the coexistence ranges
\begin{linenomath}
\begin{align*}
 \textnormal{A})&& \frac{\langle k \rangle}{i(i+3)}<z&<\frac{\langle k \rangle}{i^2},&
\\
\textnormal{B})&& \frac{\langle k \rangle}{\lambda(\mathbf{\check M})}<z&<\frac{\langle k \rangle}{\lambda(\mathbf{\hat M})}.&
\end{align*}
\end{linenomath}
In case A, we wrote the largest eigenvalues of the corresponding $\mathbf{M}$-matrices explicitly, as they assume a very simple form.
In case B, the threshold values are given through the following matrices,
\begin{linenomath}
\begin{align*}
 \mathbf{\check M}&=
\begin{pmatrix}
 j(j+1) & 2j\\
2i & i(i+1)
\end{pmatrix},&
\mathbf{\hat M}&=
\begin{pmatrix}
 j(j-1)&-2j\\
2i & i(i+1)
\end{pmatrix}.
\end{align*}
\end{linenomath}
Here we used that the equilibrium patch densities for a two-habitat system are given by $p_i^*=1/2-i(i+1)\bar z/2$ and $[j]=(j+1)\bar z$, for $j<i$. The change in $p_i$ compared to a single-habitat system (see Eq.\,\eqref{eqdens}) is due to normalization. 

\begin{figure}
\begin{center}
\includegraphics[width=.8\textwidth,trim=7cm 8cm 3cm 11cm]{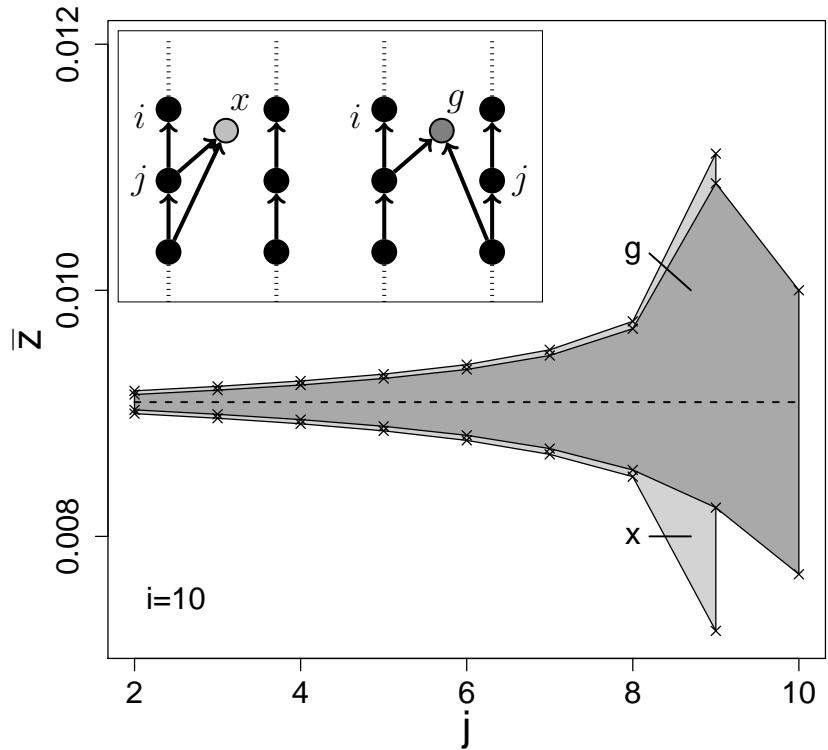}
\caption{Comparison between the coexistence ranges for omnivores and generalists. For a system of two types of habitat and 10 trophic levels (20 specialists in two branches), we plot the coexistence range for an omnivore $x$, feeding upon prey species from exclusively one branch and the coexistence range for a generalist $g$, feeding upon two prey species from both branches (inset). Throughout the whole $j$-range, the coexistence range for the omnivore is larger than the coexistence range for the generalist.}
\label{compbranches}
\end{center}
\end{figure}

Analyzing the dependence of the coexistence ranges for generalists with respect to the trophic levels of their prey, we find qualitatively the same characteristics as for omnivores:
The coexistence range increases with decreasing trophic levels and with decreasing distance between the levels. 
One exception is the case where both prey levels are equal ($i=j$). Here, the coexistence range is smaller than in the asymmetric case $i=j+1$, though still larger than for $i>j+1$ (see Fig. \ref{compbranches}).
Thus, as above, the maximal coexistence range for generalists is found when feeding on adjacent trophic levels.

In order to relate the coexistence range to the omnivores discussed above, we have to determine the coexistence range for omnivores in the presence of a second independent chain.
One finds that the coexistence range for omnivores, given in its general form in Eqs.\,\eqref{gencoex1}, Eq.\,\eqref{gencoex2}, aquires an additional factor of $1/2$ in every threshold (see Appendix).
Then, a comparison shows that the coexistence range for omnivores is always (slightly) larger than the corresponding one for generalists considered in this section (see Fig.~\ref{compbranches}).

Using the conservation equation \eqref{rel}, one can obtain an analogous result for the patch density of generalists, as the one provided in Eq.\,\eqref{pxeq} for omnivores.

\section{Summary and outlook}
\label{Sum}
In this paper we studied extensions of the metacommunity model proposed in \cite{pillai2011}.
We developed a mathematical method for the computation of persistence and coexistence conditions for different species, based on a successive linear stability analysis of the system in equilibrium. 
This method allowed convenient computation of the range of connectivities of the underlying spatial network, for which competing species can coexists in different scenarios.  

Among other results we used the proposed method to show that that the coexistence range is increased   
when a) predators feed on low trophic levels and b) when the distance between the trophic levels of their prey is small. 
This was found to be true both for predators feeding on species from the same chain of specialists and predators feeding on prey from different chains of specialists.   

For simplicity, we considered only species which are capable of feeding at most on two prey species. 
We emphasize, the mathematical method proposed here can be directly extended to predators feeding on a larger number of prey species. 

While the underlying model makes strong simplifying assumptions, the results obtained here and in the previous works \cite{pillai2010,pillai2011,pillai2012} seem plausible and thus probably merit further investigation in the future. 
A clear advantage of the simple model is that it is analytically tractable and thus offers an alternative to numerical investigations. This advantage is reinforced by the method proposed here, which allows direct computation of persistence conditions even in complex scenarios. 

A possible extension not discussed here is to consider directed geographical networks. For instance for dispersal in river networks the directed flow between patches should not be neglected, which leads to a directed connectivity \cite{muneepeerakul2007}. In this case a number of adjustments have to be made, for instance replacing the mean degree by the mean in-degree in the equations.
Another promising direction for future research is to extend the present work to strongly heterogeneous networks, such as scale-free networks, where the degree distribution cannot be approximated by the mean degree

Finally, we note that the method proposed here may also be applicable in other systems. For example, the model discussed here is closely related to ongoing investigations on simultaneous spreading of several (competing) diseases in social networks \cite{karrer2011,newman2005,masuda2006,ahn2006}.

\newpage
\section*{Appendix A: Systems with multiple types of habitat}
Starting from an empty patch network with equal distribution of $h_1$- and $h_2$-patches, specialists feeding on one or the other resource will arrange in two independent chains. 
The equilibrium patch-densities for empty habitat and $l$ specialists feeding upon each other in each of the two chains are then given by
\begin{linenomath}
\begin{align*}
\begin{split}
[h_1]^*=[h_2]^*&=\bar z\\
[i]^*&=(i+1)\bar z,  \hspace{2.1cm}1\le i < l\\
[l]^*&=\tfrac{1}{2}-\tfrac{1}{2}l(l+1)\bar z  \\
p_i^*&=\tfrac{1}{2}-\tfrac{1}{2}i(i+1)\bar z, \hspace{1cm}1\le i \le l
\end{split}
\end{align*}
\end{linenomath}
Following the same lines as in the case of one habitat type, the invasion condition for a specialist $l$ becomes
\begin{linenomath}
\begin{equation*}
z<\frac{\langle k \rangle}{l(l+1)}.
\end{equation*}
\end{linenomath}
Comparing this result to Eq.\,\eqref{specinv}, we find that the invasion threshold is divided by two, due to the presence of two independent specialist chains.
It is straight forward to verify that in general the existence of $m$ different, equally distributed habitat types leads to the formation of $m$ linear chains where the inequality
\begin{linenomath}
\begin{equation*}
z<\frac{2\langle k \rangle}{m l(l+1)}.
\end{equation*}
\end{linenomath}
determines the invasion threshold for specialist species at trophic level $l$. Consequently, the patch density $p_i$ is then given by $p_i^*=1/m-i(i+1)\bar z/2$.
Analogously, it can be shown that the persistence thresholds for omnivores and generalists acquire an additional factor $1/m$.

\section*{Appendix B: Equilibrium patch density of omnivores}
The evolution equation for the patch density $p_x$ of an omnivore $x$ feeding on level $j-1$ and $i-1$ is given by 
\begin{linenomath}
\begin{equation}
 \label{genev}
\tfrac{d}{dt}
\vec{p}_x
=\mathbf{S}_x\vec{p}_x,
\end{equation}
\end{linenomath}
where $\mathbf{S}_x$ denotes the transition matrix for $x$-patches away from the $x$-free equilibrium and $\vec{p}_x=([(j-1)x],[(i-1)x])$.
For example, in case II ($i>j+1$), the corresponding transition matrix is given by
\begin{linenomath}
\begin{align*}
\mathbf{S}_x&=
\begin{pmatrix}
 [j-1]-p_j-j\bar z & [j-1]+\bar z\\
&\\
 [i-1] & [i-1]-p_i-i\bar z
\end{pmatrix}.
\end{align*}
\end{linenomath}
In contrast to the matrix in Eq.\,\eqref{smatrix2},  describing the $x$-free equilibrium,  the amount of suitable resource patches form omnivore $x$ is reduced, as patches where the omnivore is already established are not available. However, with Eq.\,\eqref{rel} we can relate the transition matrix $\mathbf{S}_x$ to the corresponding matrix $\mathbf{S^*}$ of the $x$-free case:
\[
\mathbf{S}_x=
\mathbf{S^*}-\begin{pmatrix}
                    [(j-1)x] & [(j-1)x]\\
	           [(i-1)x] & [(i-1)x]
                    \end{pmatrix}.
\]
Rewriting Eq.\,\eqref{genev} reveals,
\begin{linenomath}
\begin{align*}
\tfrac{d}{dt}
\vec{p}_x
&=\Bigl\{\mathbf{S^*}-\begin{pmatrix}
                    [(j-1)x] & [(j-1)x]\\
	           [(i-1)x] & [(i-1)x]
                    \end{pmatrix}\Bigr\}
\vec{p}_x  =     (\mathbf{S^*}-p_x)\vec{p}_x.
\end{align*}
\end{linenomath}
So, at equilibrium (with $x$), the density of $x$-patches is given by
\begin{linenomath}
 \begin{equation*}
  p_x^*=[(j-1)x]^*+[(i-1)x]^*=\lambda(\mathbf{S^*}).
\end{equation*}
\end{linenomath}

\section*{Appendix C: Example for omnivory in a $4$-trophic chain}

\begin{figure}
\begin{center}
\includegraphics[width=.9\textwidth, trim= 5cm 16.5cm 4cm 7cm]{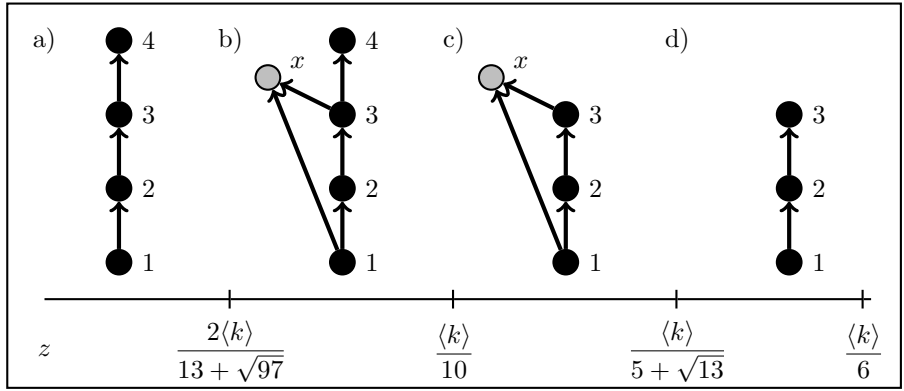}
\caption{Example for a maximal food web emerging in the metacommunity. In contrast to Fig.~\ref{singlebr}, trophic levels of the omnivores' prey species differ by more than one. Symbols are the same as in Fig.~\ref{singlebr}. Shown are different food web configurations which arise in the given parameter ranges. The corresponding parameter ranges (not true to scale) are obtained by the mathematical formalism presented in the text. }
\label{singlebr2}
\end{center}
\end{figure}

Here, we consider an additional example to the one provided in the main text, in order to demostrate the procedure stated in Fig.~\ref{recipebox}.
In this example, an omnivore feeds on two prey species which are not at subsequent trophic levels (see Fig.~\ref{singlebr2}). 
Consequently, for step 1) we have to use the $\mathbf{S}^*$-matrix given in \eqref{smatrix2} for $i=4$ and $j=2$,
\begin{linenomath}
\begin{equation*}
\mathbf{S^*}=
 \begin{pmatrix}
 \langle k \rangle ([1]^*-p_2^*)-2 z &  \langle k \rangle[1]^*+ z\\&\\
 \langle k \rangle[3]^* &  \langle k \rangle([3]^*-p_4^*)-4 z
 \end{pmatrix}.
\end{equation*}
\end{linenomath}
Following step 2) and step 3), we construct 
\begin{linenomath}
\begin{align*}
\mathbf{\check{S}^*}&=
\underbrace{
\begin{pmatrix}
 -1 &0\\0 & -1
\end{pmatrix}}_{\mathbf{\check{s}_1}} \langle k \rangle
+
\underbrace{
\begin{pmatrix}
 3 &3\\4 & 10
\end{pmatrix} }_{\mathbf{\check{s}_2}}z,&p_4^*\neq0,\\
\mathbf{\hat{S}^*}&=
\underbrace{
\begin{pmatrix}
 -1 &0\\1 & 1
\end{pmatrix}
}_{\mathbf{s_1}} \langle k \rangle
+
\underbrace{
\begin{pmatrix}
 3 &3\\-6 & -10
\end{pmatrix}
}_{\mathbf{s_2}}z, &p_4^*=0,
\end{align*}
\end{linenomath}
using the equilibrium patch densities in Eq.\,\eqref{eqdens}.
Then, in step 4), we determine the matrices
\begin{linenomath}
\begin{equation*}
\mathbf{\check{M}}=
\underbrace{
\begin{pmatrix}
 3 &3\\4 & 10
\end{pmatrix} }_{\mathbf{\check{s}_2}}\cdot
\underbrace{
\begin{pmatrix}
 1 &0\\0 & 1
\end{pmatrix}}_{-\mathbf{\check{s}_1}^{-1}}=
 \begin{pmatrix}
  3&3\\4&10
 \end{pmatrix}
\end{equation*}
\end{linenomath}
and 
\begin{linenomath}
\begin{equation*}
\mathbf{\hat{M}}=
\underbrace{
\begin{pmatrix}
 3 &3\\-6 & -10
\end{pmatrix}
}_{\mathbf{s_2}}
\cdot
{\underbrace{
\begin{pmatrix}
 1 &0\\-1 & -1
\end{pmatrix}}_{-\mathbf{s_1}^{-1}}}
=\begin{pmatrix}0&-3\\4&10\end{pmatrix},
\end{equation*}
\end{linenomath}
and calculate their largest eigenvalues:
\[
\lambda(\mathbf{\check{M}})=13+\sqrt{97}, \quad \quad \lambda(\mathbf{\hat{M}})=5+\sqrt{13}.
\]
According to step 5), the parameter range where an omnivore feeding on trophic levels 1 and 3 can coexist with a specialist chain is given by
\begin{linenomath}
\begin{equation*}
 \frac{\langle k \rangle}{13+\sqrt{97}}<z<\frac{\langle k \rangle}{5+\sqrt{13}}.
\end{equation*}
\end{linenomath}
Comparing this range to the one given in Eq.\,\eqref{coexx}, one can observe that the whole range is shifted to smaller $z$-values, and that the total size of the coexistence range is smaller compared to the previous example. 
These two features resemble the general behavior shown in Fig.~\ref{ijdep}.

\begin{figure}[tb]
\begin{center}
\includegraphics[width=.6\textwidth]{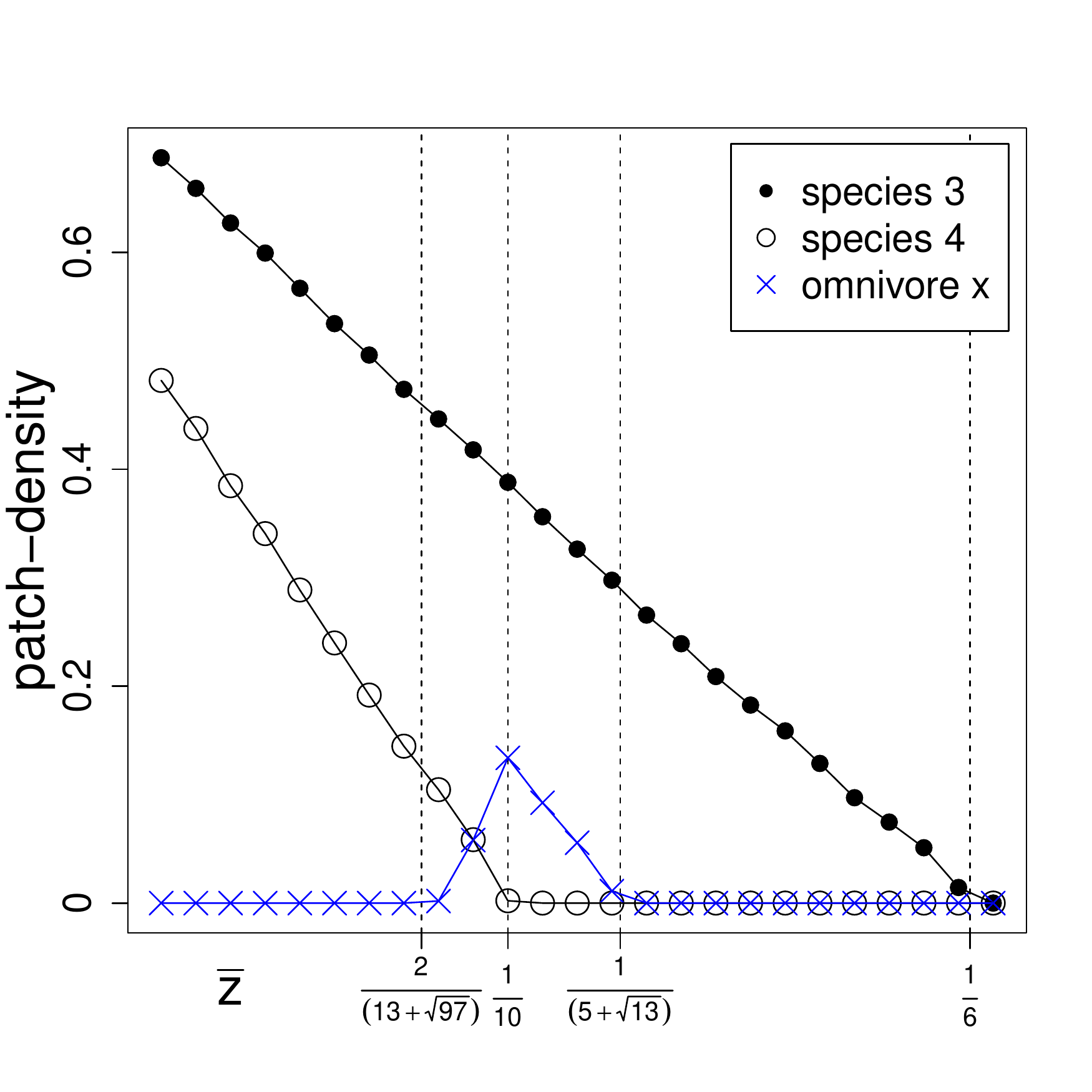}
\caption{Comparison of the predicted ranges with simulation results. Plotted are the equilibrium patch-densities $p_3^*$ ($\bullet$), $p_4^*$ ($\circ$) and $p_x^*$($\times$), obtained from network simulations. Eeach point corresponds to an average over 10 independent simulation runs. We used $N=10000, \langle k \rangle=20, c=0.05$ and varied $\bar z=e$ within the indicated range. Dashed lines correspond to the persistence thresholds predicted by our analytical calculations.
Coexistence range for omnivore $x$: $2/(13+\sqrt{97})<\bar z<1/(5+\sqrt{13})$, persistence threshold for specialist 3: $\bar z=1/6$ and persistence threshold for specialist 4: $\bar z=1/10$.}
\label{sim_coex}
\end{center}
\end{figure}

A summary of the obtained results for this example system can be found in Fig.~\ref{singlebr2}.
In Fig.~\ref{sim_coex} we show simulation results for this system. It can be seen that the analytically estimated ranges correspond to disappearance or appearance of the respective
species types, indicating that the predicted food web configurations actually are realized in the metacommunity.


\newpage

\end{document}